\overfullrule=0pt
\input harvmac

\def\a{\alpha}
\def\b{\beta}
\def\g{\gamma}
\def\d{\delta}
\def\r{\rho}
\def\s{\sigma}
\def\l{\lambda}
\def\o{\omega}
\def\Pb{\overline\Pi}
\def\A{\cal A}

\def\N{\nabla}
\def\Nb{\overline\N}
\def\Jb{\overline J}
\def\p{\partial}
\def\pb{\overline\p}
\def\O{\Omega}
\def\Ab{\overline A}
\def\Ob{\overline\O}
\def\S{\Sigma}
\def\ve{\varepsilon}
\def\Os{\O^{(s)}}
\def\ws{\o^{(CS)}}
\def\L{\Lambda}
\def\H{\widehat H}
\def\wh{\widehat\o}
\def\G{\Gamma}

\Title{ \vbox{\baselineskip12pt \hbox{DFPD 07/TH/13 } \hbox{} }}
{\vbox{\centerline{ BRST Anomaly and Superspace Constraints of the }
\smallskip
\centerline{ Pure Spinor Heterotic String in a Curved Background }
}}

\smallskip
\centerline{Osvaldo Chand\'{\i}a\foot{e-mail: ochandia@unab.cl}}
\smallskip
\centerline{\it Departamento de Ciencias F\'{\i}sicas, Universidad
Andr\'es Bello} \centerline{\it Rep\'ublica 252, Santiago, Chile}
\bigskip
\centerline{Mario Tonin\foot{e-mail: tonin@pd.infn.it}}
\smallskip
\centerline{\it Dipartimento de Fisica, Universit\`a degli Studi di
Padova} \centerline{\it INFN Sezione di Padova, Italia}

\bigskip

\bigskip

\noindent The pure spinor heterotic string in a generic super
Yang-Mills and supergravity background is considered. We determine
the one-loop BRST anomaly at the cohomological level. We prove that
it can be absorbed by consistent corrections of the classical
constraints due to Berkovits and Howe, in agreement with the
Green-Schwarz cancelation mechanism.

\Date{July 2007}


\newsec{Introduction}

The pure spinor formalism of superstrings
\ref\BerkovitsFE{N.~Berkovits, ``Super-Poincare Covariant
Quantization of the Superstring,'' JHEP 0004 (2000) 018
[arXiv:hep-th/0001035]\semi N.~Berkovits, ``Relating the RNS and
Pure Spinor Formalisms for the Superstring,'' JHEP 0108 (2001) 026
(2001) [arXiv:hep-th/0104247]\semi N.~Berkovits, ``ICTP Lectures on
Covariant Quantization of the Superstring,''
arXiv:hep-th/0209059\semi N.~Berkovits, ``Multiloop Amplitudes and
Vanishing Theorems Using the Pure Spinor Formalism for the
Superstring,'' JHEP 0409 (2004) 047 [arXiv:hep-th/0406055]\semi
N.~Berkovits and B.~C.~Vallilo, ``Consistency of Super-Poincare
Covariant Superstring Tree Amplitudes,'' JHEP 0007 (2000) 015
[arXiv:hep-th/0004171]\semi N.~Berkovits and O.~Chand\'{\i}a,
``Lorentz Invariance of the Pure Spinor BRST Cohomology for the
Superstring,'' Phys.\ Lett.\  B514 (2001) 394
[arXiv:hep-th/0105149].} and the Green-Schwarz (GS) formalism
\ref\GreenWT{M.~B.~Green and J.~H.~Schwarz, ``Covariant Description
of Superstrings,'' Phys.\ Lett.\  B136 (1984) 367.} are synoptical,
in  that both are based on the embedding of the superstring
world-sheet into the target superspace and have manifest
supersymmetry. The deep relation between the two approaches is
discussed in \ref\OdaZM{I.~Oda and M.~Tonin, ``On the Berkovits
Covariant Quantization of GS Superstring,'' Phys.\ Lett.\  B520
(2001) 398 [arXiv:hep-th/0109051]\semi M.~Matone, L.~Mazzucato,
I.~Oda, D.~Sorokin and M.~Tonin, ``The Superembedding Origin of the
Berkovits Pure Spinor Covariant Quantization of Superstrings,''
Nucl.\ Phys.\  B639 (2002) 182 [arXiv:hep-th/0206104].},
\ref\AisakaVN{Y.~Aisaka and Y.~Kazama, ``Origin of Pure Spinor
Superstring,'' JHEP 0505 (2005) 046 [arXiv:hep-th/0502208].},
\ref\BerkovitsTW{N.~Berkovits and D.~Z.~Marchioro, ``Relating the
Green-Schwarz and Pure Spinor Formalisms for the Superstring,'' JHEP
0501 (2005) 018 [arXiv:hep-th/0412198].}. The pure spinor approach
has the great advantage over the GS one to allow for a consistent
and covariant quantization of the superstring. The GS approach
cannot be quantized covariantly due to its peculiar
$\kappa$-symmetry \ref\SiegelHH{W.~Siegel, ``Hidden Local
Supersymmetry in the Supersymmetric Particle Action,'' Phys.\ Lett.\
B128 (1983) 397.} which cannot be gauge fixed in a Lorentz covariant
way. In the pure spinor formulation the $\kappa$-symmetry is
replaced by a symmetry generated by the BRST charge $Q = \oint \l^\a
d_\a$ where $ \l^\a $ is a pure spinor.

One of the main applications of the pure spinor formalism is the
construction of string actions on supersymmetric backgrounds
\ref\BerkovitsUE{N.~Berkovits and P.~S.~Howe, ``Ten-dimensional
Supergravity Constraints from the Pure Spinor Formalism for the
Superstring,'' Nucl.\ Phys.\  B635 (2002) 75
[arXiv:hep-th/0112160].}, \ref\ChandiaIX{O.~Chand\'{\i}a, ``A Note
on the Classical BRST Symmetry of the Pure Spinor String in a Curved
Background,'' JHEP 0607 (2006) 019 [arXiv:hep-th/0604115].}, \OdaZM,
including those with Ramond-Ramond fields like anti de Sitter
space-times \ref\BerkovitsYR{N.~Berkovits and O.~Chand\'{\i}a,
``Superstring Vertex Operators in an AdS$_5\times$S$^5$
Background,'' Nucl.\ Phys.\  B596 (2001) 185
[arXiv:hep-th/0009168]\semi B.~C.~Vallilo, ``Flat Currents in the
Classical AdS$_5\times$S$^5$ Pure Spinor Superstring,'' JHEP 0403
(2004) 037 [arXiv:hep-th/0307018]\semi N.~Berkovits, ``Quantum
Consistency of the Superstring in AdS$_5\times$S$^5$ Background,''
JHEP 0503 (2005) 041 [arXiv:hep-th/0411170]\semi N.~Berkovits, ``A
New Limit of the AdS$_5\times$S$^5$ Sigma Model,''
arXiv:hep-th/0703282.}. A common feature of both approaches in
curved backgrounds is that, in the associated $\s$-models, the
requirement of invariance and nilpotence under $\kappa$-symmetry in
one case \ref\WittenNT{E.~Witten, ``Twistor - Like Transform in
Ten-dimensions,'' Nucl.\ Phys.\  B266 (1986) 245.} and under the
BRST symmetry in the other case \BerkovitsUE, \ChandiaIX, implies,
to zero order in $\a'$, constraints for the background torsion and
curvatures that force the background fields to be on shell.

A relevant question in this context is to understand and compute the
corrections to these constraints to higher order in $\a'$.

There are two ways to study these corrections for the GS
superstring. The first method computes the relevant $\b$-functions
and imposes that they vanish to reach conformal invariance at the
quantum level. The vanishing of the $\b$-functions determines the
corrections to the background field equations
\ref\GrisaruSA{M.~T.~Grisaru, H.~Nishino and D.~Zanon, ``Beta
Functions for the Green-Schwarz Superstring,'' Nucl.\ Phys.\  B314
(1989) 363\semi M.~T.~Grisaru and D.~Zanon, ``The Green-Schwarz
Superstring Sigma Model,'' Nucl.\ Phys.\  B310 (1988) 57\semi
P.~E.~Haagensen, ``The Lorentz-Chern-Simons form in Heterotic
Green-Schwarz Sigma Models,'' Mod.\ Phys.\ Lett.\  A6 (1991) 431
(1991).}. The second method is cohomological in nature: it
classifies and then computes the anomalies of the BRST
$\kappa$-symmetry. These anomalies determine the $\a'$ corrections
of the torsion and curvatures constraints \ref\AtickIY{J.~J.~Atick,
A.~Dhar and B.~Ratra, ``Superstring Propagation in Curved Superspace
in the Presence of Background Super Yang-Mills Fields,'' Phys.\
Lett.\  B169 (1986) 54.}, \ref\ToninIT{M.~Tonin, ``Superstrings, k
Symmetry and Superspace Constraints,'' Int.\ J.\ Mod.\ Phys.\  A3
(1988) 1519\semi ``Consistency Condition for kappa Anomalies and
Superspace Constraints in Quantum Heterotic Superstrings,'' Int.\
J.\ Mod.\ Phys.\  A4 (1989) 1983\semi ``Covariant Quantization and
Anomalies of the G-S Heterotic Sigma Model,'' Int.\ J.\ Mod.\ Phys.\
A6 (1991) 315.}, \ref\CandielloCD{A.~Candiello, K.~Lechner and
M.~Tonin, ``k Anomalies and Space-time Supersymmetry in the
Green-Schwarz Heterotic Superstring,'' Nucl.\ Phys.\  B438 (1995) 67
[arXiv:hep-th/9409107].}. The equivalence of the two methods becomes
clear if one notices that the square of the $\kappa$-symmetry
transformations produces a Weyl-Lorentz (i.e. conformal) world-sheet
transformation. It is remarkable that correction of order $n$ in
$\a'$ in the conformal approach appear at  order $(n-1)$ in the
cohomological approach.

Since the pure spinor formulation describes a critical string, one
expects that the conformal invariance is preserved on shell also at
the quantum level and the $\b$-functions vanish
\ref\ChandiaHN{O.~Chand\'{\i}a and B.~C.~Vallilo, ``Conformal
Invariance of the Pure Spinor Superstring in a Curved Background,''
JHEP 0404 (2004) 041 [arXiv:hep-th/0401226].} \ref\BedoyaIC{
O.~A.~Bedoya and O.~Chand\'{\i}a, ``One-loop Conformal Invariance of
the Type II Pure Spinor Superstring in a Curved Background,'' JHEP
0701 (2007) 042 [arXiv:hep-th/0609161].}. However, as discussed in
\BerkovitsUE, the vanishing of the $\b$-function is not sufficient
to determine the corrections of the field equations. It turns out
that the holomorphicity of the BRST current and the nilpotence of
the BRST charge are also needed. Equivalently one can apply the
cohomological method and study the anomalies of the BRST symmetry
generated by $Q$.

For the heterotic string, both in the GS and in the pure spinor
approaches, the constraints that arise at zero order in $\a'$
describe a model where the B-field is decoupled from the gauge
sector. Then, at first order in $\a'$, one expects a correction
related to the gauge and Lorentz Chern-Simons three-form, in order
to cancel the gauge and Lorentz anomalies by the standard
Green-Schwarz mechanism \ref\HullJV{ C.~M.~Hull and E.~Witten,
``Supersymmetric Sigma Models and the Heterotic String,'' Phys.\
Lett.\  B160 (1985) 398.}.

In the GS formulation, this correction was indeed found, as an
anomaly of the $\kappa$-symmetry, for the Yang-Mills Chern-Simons
form in \AtickIY\ and for the full (gauge and Lorentz) Chern-Simons
form in \ToninIT. The coefficients of this anomaly has been
explicitly computed in \CandielloCD, in agreement with the GS
anomaly cancelation mechanism. One should notice that in order to
implement the consistency condition for the Lorentz anomaly, a
theorem  to obtain a solutions of the SUGRA-SYM constraints in
presence of the gauge and Lorentz Chern-Simons forms has to be used
\ref\BonoraIX{L.~Bonora, P.~Pasti and M.~Tonin, ``Superspace
Formulation of 10-D Sugra+Sym Theory a la Green-Schwarz,'' Phys.\
Lett.\  B188 (1987) 335.}, \ref\BonoraXN{L.~Bonora, M.~Bregola,
K.~Lechner, P.~Pasti and M.~Tonin, ``Anomaly Free Supergravity and
Super Yang-Mills Theories in Ten-dimensions,'' Nucl.\ Phys.\  B296
(1988) 877.}, \ref\D'AuriaPC{R.~D'Auria and P.~Fre, ``Minimal 10-D
Anomaly Free Supergravity and the Effective Superstring Theory,''
Phys.\ Lett.\ B200 (1988) 63\semi R.~D'Auria, P.~Fre, M.~Raciti and
F.~Riva, ``Anomaly Free Supergravity in D = 10. 1. The Bianchi
Identities and the Bosonic Lagrangian,'' Int.\ J.\ Mod.\ Phys.\  A3
(1988) 953\semi L.~Castellani, R.~D'Auria and P.~Fre, ``What we
Learn on the Heterotic String Vacua from Anomaly Free
Supergravity,'' Phys.\ Lett.\  B196 (1987) 349\semi M.~Raciti,
F.~Riva and D.~Zanon ``Perturbative approach to D=10 superspace
supergravity with a Lorentz-Chern-Simons form'' Phys. \ Lett. B227
(1989) 118 .}.\foot{ A procedure to obtain corrections to SUGRA/SYM
system order by order in $\a'$ was done in \ref\BellucciFA{
S.~Bellucci, D.~A.~Depireux and S.~J.~J.~Gates, ``Consistent and
Universal Inclusion of the Lorentz Chern-Simons Form in D = 10, N =
1 Supergravity Theories,'' Phys.\ Lett.\  B238 (1990) 315.}.
Unfortunately, this approach, as developed in \BellucciFA, leads to
inconsistencies (see \ref\BonoraTX{ L.~Bonora {\it et al.}, ``Some
Remarks on the Supersymmetrization of the Lorentz Chern-Simons Form
in D = 10 N = 1 Supergravity Theories,'' Phys.\ Lett.\  B277 (1992)
306.}).}

In this paper we consider the problem of determining the $\a'$
corrections of the heterotic string $\s$-model, in the framework of
the pure spinor approach, looking for the BRST anomalies at the
cohomological level. In particular we shall obtain  the full
expression of the anomaly related to the gauge and Lorentz
Chern-Simons three-form, which arises at first order in $\a'$.

In the next section we will review the pure spinor construction for
the heterotic string in a generic SYM/SUGRA background. In section 3
we determine the form in which the theorem of \BonoraIX\ is
implemented with the constraints for background fields of
\BerkovitsUE. In section 4 we propose a one-loop anomaly for the
BRST symmetry and show that it is cohomologically non trivial.
Finally, we end with a conclusion section. Before finish this
section, we shall introduce our notation.

\subsec{Notation}

Our normalization for $n$-(super)forms is

$$ F = {1\over{n!}} dZ^{M_1}...dZ^{M_n}
F_{M_{n}...M_{1}} = {1\over{n!}} E^{A_1}...E^{A_n}F_{A_{n}...A_{1}}
,$$ where $Z^M$ are the ten dimensional $N=1$ superspace
coordinates, $E^{A} = dZ^{M} E_{M}{}^{A}$ are the supervielbeins. We
use latin letters for vector-like indices, greek letters for
spinor-like indices and Capital letters for both. Letters from the
beginning of the alphabet denote flat (Lorentz) indices and letters
from the middle of alphabet are for curved ones. Once a set of
supervielbeins is specified, an $n$ superform can be decomposed as

$$F = \sum F_{p,q} ,$$ where $F_{p,q}$ denote the
component of $F$ with $p$ vector-like vielbeins and $q = n - p$
spinor-like vielbeins.

\newsec{The Heterotic String Action}

The sigma model action for the heterotic string in a SUGRA/SYM
background in the pure spinor formalism is given by \BerkovitsUE\

\eqn\action{ S = {1\over\a'} \int d^2z ~[ \ha \Pi^a \Pb^b \eta_{ab}
+ \ha \Pi^A \Pb^B B_{BA} + \o_\a \Nb \l^\a }
 $$
 + d_\a ( \Pb^\a + \Jb^I W_I^\a ) + \Pi^A A_{IA}\Jb^I + \l^\a \o_\b \Jb^I
U_{I\a}{}^\b ] + S_J  + S_{FT},$$ where $( \Pi^A, \Pb^A ) = ( \p Z^M
E_M{}^A, \pb Z^M E_M{}^A )$, $\l^\a$ is a pure spinor and $ \o_\a$
is its conjugate momentum. The covariant derivative for the pure
spinor $\l^\a$ is given by $\Nb \l^\a = \pb \l^\a + \l^\b \pb Z^M
\O_{M\b}{}^\a$, where $\O_{M\a}{}^\b$ is the connection for the
structure group and it has the form $\O_{M\a}{}^\b = \Os_M
\d_\a{}^\b + {1\over 4} \O_{Mab} (\g^{ab})_\a{}^\b$. The world-sheet
field $d_\a$ has conformal weight $(1,0)$ and plays the role of
generating translations in superspace. $\Jb^I$ ($I = 1,\dots, 496$),
with conformal weight $(0,1)$, are the currents of the gauge group,
$SO(32)$ or $E_8 \times E_8$ and $dZ^M A_{IM}$ is the gauge group
connection. $S_J$ is the free action for the heterotic fermions. The
superfield $W^\a_I$ has the gaugino as the lowest component and
$U_{I\a}{}^\b$ contains the field strength for the gauge boson in
its lowest component. Finally, $S_{FT}$ is the Fradkin-Tseytlin term
given by

\eqn\sft{ S_{FT} = \int d^2z ~ r^{(2)} \Phi ,} where $r^{(2)}$ is
the world-sheet curvature and $\Phi$ is the dilaton superfield.
Although the Fradkin-Tseytlin term breaks the classical conformal
invariance of the action \action, it helps to restore it at the
quantum level as it was shown in \ChandiaHN\ in the one-loop case.
Note that the dilaton superfield is related to the Weyl part of the
curvature connection as $\N_\a \Phi = 4 \Os_\a$.

Besides the action \action\ being classically invariant under
conformal transformations, it is invariant under gauge
transformations and a pair of Lorentz transformations acting on the
background fields. These two Lorentz transformations act
independently on the bosonic local indices, e.g. $\d \Pi^a = \Pi^b
\S_b{}^a$, and on the fermionic local indices, e.g. $\d d_\a = -
\S_\a{}^\b d_\b$. Both Lorentz transformations can be identified as
it is done in \BerkovitsUE.

The pure spinor superstring  has a very important symmetry, it is
invariant under the BRST-like pure spinor transformation
\BerkovitsFE\ generated by the pure spinor BRST charge $Q = \oint
\l^\a d_\a$. As it was stressed in \BerkovitsUE, one must demand
that also the action \action\ is invariant  under this  symmetry. By
demanding nilpotence and world-sheet time conservation of $Q$, the
action \action\ turns out to be invariant if the background
superfields satisfy suitable constraints which determine the
SUGRA/SYM equations of motion for them. Nilpotence is achieved by
demanding \eqn\nilp{\l^\a \l^\b T_{\a\b}{}^A = \l^\a \l^\b H_{\a\b
A} = \l^\a \l^\b F_{I\a\b} = \l^\a \l^\b \l^\g R_{\a\b\g}{}^\d  =
0,} where $T^A$  is the torsion 2-form, $H = dB$, $F$ is the
field-strength two form and $R_\a{}^\b$ is the curvature two form.
We use the notation of \ChandiaHN. The charge conservation can be
obtained by determining the equations of motion of \action\ and then
imposing $\pb ( \l^\a d_\a ) = 0$ \BerkovitsUE\ or by demanding
invariance of \action\ under the BRST transformations \ChandiaIX. In
this case, the action transforms as

\eqn\qaction{Q S = {1\over\a'} \int d^2z ~ [\ha \l^\a \Pi^a \Pb^b (
T_{\a(ab)} + H_{ba\a} ) + \ha \l^\a \Pi^\b \Pb^a ( H_{\b\a a} -
T_{\b\a a} ) + \l^\a d_\b \Pb^a T_{a\a}{}^\b }
$$
- \l^\a \l^\b d_\g \Pb^a R_{a\a\b}{}^\g + \l^\a \Pi^a \Jb^I ( \ha (
H_{\a\b a} + T_{\a\b a} ) W_I^\b - F_{Ia\a} ) + \l^\a \Pi^\b \Jb^I (
\ha H_{\a\b\g} W_I^\g - F_{I\a\b} )
$$
$$
+ \l^\a d_\b \Jb^I ( U_{I\a}{}^\b - W_I^\g T_{\g\a}{}^\b - \N_\a
W_I^\b ) + \l^\a \l^\b \o_\g ( \N_\a U_{I\a}{}^\g + W_I^\d
R_{\d\a\b}{}^\g )].$$ As it was shown in \BerkovitsUE, the
nilpotence constraints \nilp\ and the vanishing of \qaction\ allow
to write the following constraints for the torsion and curvature
components

\eqn\torsion{ T_{a\a}{}^\b = T_{\a\b}{}^\g = 0,\quad T_{\a\b}{}^a =
\g^a_{\a\b},\quad T_{\a a}{}^b = 2(\g_a{}^b)_\a{}^\b \O_\b ,}

\eqn\curv{ H _{\a\b\g} = H_{a\b\g} - (\g_a)_{\b\g}= 0 ,}

\eqn\fab{ F_{I\a\b} = 0,} where $\g^{a}_{\a\b}$ and $(\g^a)^{\a\b}$
denote the usual Pauli matrices, i.e. the off-diagonal blocks of the
Dirac matrices, so that they are symmetric in $(\a,\b)$. Besides,
Bianchi identities imply that the torsion component $T_{abc} =
\eta_{cd} T_{ab}{}^d$ is completely antisymmetric \ChandiaHN. Note
that the torsion component $T_{\a\b}{}^\g$ can be set to zero only
after the use of the `shift' symmetry of \BerkovitsUE.

Note that in \qaction, the field equation

\eqn\pib{ \Pb^\a + \Jb^I W_I^\a = 0 ,} which follows from varying
the action \action\ respect to the world-sheet field $d_\a$, has
been used.

Finally one must  require that the action \action\ is also invariant
under the ``$\o$-symmetry'' $ \d \o_\a = (\g^a\l)_\a \L_a$, where
$\L_a$ are local parameters, which implies that

$$U_{I\a}{}^\b = U_I \d_\a{}^\b +  U_{I ab} (\g^{ab})_\a{}^\b .$$

A natural question to be addressed at this point is the quantum
preservation of the symmetries of \action. In particular, the
possibility of finding $\a'$ corrections to the constraints of
\nilp\ and those from the vanishing of \qaction. Let us first
discuss the gauge and Lorentz anomalies\foot{The following paragraph
is based on discussions with N. Berkovits and V. Pershin.} .

The anomaly for the local symmetries is

\eqn\local{ \d \G_{eff} = \int d^2 z ~ [\ha ( \pb A_I - \p \Ab_I )
\ve_I + {1\over 4} ( \pb \O_{ab} - \p \Ob_{ab} ) \S^{ab}] ,} where
$\G_{eff}$ is the effective action (i.e. the generating functional
of 1PI vertex functions), $\ve_I$ and $\S^{ab}$ are the parameters
of the gauge and the Lorentz transformations respectively and $A_I =
\p Z^M A_{IM}, \Ab_I = \pb Z^M A_{IM}, \O_{ab} = \p Z^M \O_{Mab},
\Ob_{ab} = \pb Z^M \O_{Mab}$. It is also possible the presence of
terms like

$$
\int d^2 z ~ d_\a W_I^\a \pb \ve_I,$$ which can be eliminated by
adding suitable counterterms. There is also a potential anomaly
associated to $\Os$. However, this contribution vanishes because
$\Os$ appears in the combination $\Os + \Jb^I U_I$ which is zero
on-shell \ChandiaHN.

Since the quantum theory cannot be anomalous under a local symmetry,
the expression \local\ must be canceled by the standard
Green-Schwarz mechanism \HullJV. It is done by allowing the $B$
two-form superfield not to be inert under gauge and Lorentz
transformations. It has to transform as

\eqn\dbgauge{ \d B = -\a' ( dA_I \ve_I + \ha d\O_{ab} \S^{ab} ) .}
In order to assure  gauge and Lorentz invariance of the $B$ field
strength $H$ one must define it as

\eqn\hgauge{ H = d B -{\a'\over 2} \ws ,} where $\ws$ is the
Chern-Simons three from given by

\eqn\cs{ \ws = tr ( A d A - {2\over 3} A^3 ) + \O^{ab} d \O_{ab} -
{2\over 3} \O_a{}^b \O_b{}^c \O_c{}^a ,} and satisfying

\eqn\dh{ dH = tr(FF) + R^{ab}R_{ab} .} Note that $H$ in \hgauge\ is
defined up to a gauge and Lorentz invariant three-form.

The classical constraints coming from \nilp\ and the vanishing of
\qaction\ lead to $dH=0$ and therefore have to be corrected. These
corrections arise as anomalies of the BRST symmetry generated by the
nilpotent charge $Q$, that is, if we define

\eqn\anom{ Q\G_{eff} = \a' {\A} ,} ${\A}$ is a non trivial cocycle
of the cohomology of $Q$, in the space of local functionals of ghost
number $1$. Then from the previous discussion it is expected that
${\A}$  will contain a term

\eqn\expec{{\A} = \ha \int d^2z ~ \l^\a \Pi^A \Pb^B \ws_{BA\a} +
\cdots ,} which modifies the definition of $H$ since the variation
of the term involving $B$ in the action \action\ is proportional to
$$ \int d^2 z ~ \l^\a \Pi^A \Pb^B ( d B )_{BA\a} .$$ In the next sections we
will determine the complete form of \expec\ by studying the
conditions coming from $Q {\A} = 0$.

\newsec{The Cohomology and an Useful Theorem}

Let us start from \expec\ and compute its BRST variation. Any
variation of the Chern-Simons term  showed in \expec\ is

$$
\d \int d^2z ~ \l^\a \Pi^A \Pb^B \ws_{BA\a} = \int d^2z ~ \d Z^M
E_M{}^C \l^\b \Pi^A \Pb^B (d\ws)_{BA\b C},$$ because $\ws$ is a
three-form. Now we recall the BRST variation for $Z^M$ to be
\ChandiaIX\

$$
\d_{BRST} Z^M = Q Z^M = \l^\a E_\a{}^M,$$ then we obtain

\eqn\qtwo{ Q {\A} = \int d^2z ~ \ha \l^\a \l^\b \Pi^A \Pb^B (d
\ws)_{BA\a\b} + \cdots  } $$ = \int d^2z ~ \l^\a \l^\b \Pi^A \Pb^B (
F_I F_I + \ha R^{ab} R_{ab} )_{BA\a\b} + \cdots ,$$ where we have
used $d\ws = tr (F F) + R^{ab} R_{ab}$. We will fix the $\cdots$
terms to make this expression to vanish.

It follows from the constraint
$ F_{I\a\b}= 0 ,$ that the 4-superform $F_I F_I$ vanishes in the sectors
$(0,4)$ and $(1,3)$ (i.e. in the sectors with $4$ or $3$ spinor-like
local indices ). Moreover in the sector $(2,2)$ $( F_I F_I
)_{ba\a\b}$ has the following structure

\eqn\ffstr{ ( F_I F_I )_{ba\a\b} = (\g_{[a})_{\a\g}(\g_{b]})_{\b\d}
W_I^\g W_I^\d .} As it will be shown in section 4, this structure is
essential to compute the anomaly for the gauge part in \qtwo\ . The
curvature part in \qtwo\ could be treated similarly if the index
structure were the same. Unfortunately, it is not the case with the
constraints of \nilp\ and \qaction. However there exists the
following result. It was shown in \ToninIT, \CandielloCD, \BonoraXN\
that with a different set of torsion constraints \BonoraIX\ (the
gauge part has the same constraints) that

\eqn\rr{R'^{ab} R'_{ab} = d X' + K,} where the three form $X'$ and
the four form $K$ are Lorentz invariant. They were determined in
\BonoraXN. The main property in \rr\ is that the four form $K$
vanishes in the sectors $(0,4)$ and $(1,3)$ and that in the sector
$(2,2)$ has the same structure than $( F_I F_I )_{ab\a\b}$, that is

$$ K_{ab\a\b} = (\g_{[a})_{\a\g} (\g_{b]})_{\b\d} K^{\g\d} ,$$ with $ K^{\g\d}
= - K^{\d\g}$. This property will be crucial to determine also the
Lorentz part in the BRST anomaly.

In order to use this result we should relate the Berkovits-Howe
constraints to the ones of \BonoraXN. Now it will be shown that
there exists a redefinition of fields which makes the job. We
redefine the vielbein one-form as

\eqn\redefvielbein{ E'^a = e^{-{1\over 3}\Phi} E^a, \quad E'^\a =
e^{-{1\over 6}\Phi} ( E^\a + {1\over 3} E^a \g_a^{\a\b} \N_\b \Phi )
,} and the components of the Lorentz superspace connection one form
as

\eqn\redcon{ \O'_{ab} = \O_{ab} +  \L_{ab} ,} where

\eqn\redom { \L_{ab} = {1\over 3} E_{[a} \N_{b]} \Phi  - {1\over 12}
E^c \g_{cab}^{\a\b} \N_\a \N_\b \Phi - {1\over 6} E^\a
(\g_{ab})_\a{}^\b \N_\b \Phi.} In components, these transformations
imply the following torsion constraints

\eqn\tonin{ T'_{\a\b}{}^a = \g^a_{\a\b},\quad T'_{\a\b}{}^\g =
T'_{\a a}{}^b = 0,\quad T'_{a\a}{}^\b = {1\over 3} e^{{1\over
6}\Phi}  (\g_a \g^{bcd} )_\a{}^\b \tau_{bcd},} where

$$ \tau_{bcd} = {1\over 96} \g_{bcd}^{\g\d} ( \N_\g \N_\d \Phi +
{4\over 3} (\N_\g \Phi) (\N_\d \Phi) ) ,$$ which correspond to a set
of constraints used in \BonoraXN\ to show the theorem \rr.

Now we can use \redcon\ to rewrite \rr\ for the Berkovits-Howe
constraints. In fact,

\eqn\RR{ R^{ab} R_{ab} = R'^{ab} R'_{ab} - d ( 2 R_{ab}\L^{ab} +
\L_{ab} \N \L^{ab} ) ,} Therefore, we have shown that
\eqn\nth{R^{ab} R_{ab} = d X + K,} where

$$X = X' - 2 R_{ab}\L^{ab} - \L_{ab} \N \L^{ab}.$$

\newsec{Quantum BRST Invariance}

From the discussion of the previous section it is expected that the
anomaly is

\eqn\anm{ {\A} = \ha \int d^2z ~ \l^\a \Pi^A \Pb^B \wh_{BA\a} +
\cdots , } where

\eqn\defwh{ \wh = \ws - X ,} and

$$d \wh = L ,$$ where the closed 4-superform $L$ is

\eqn\defl{ L = 2 F_I  F_I + R_{ab} R^{ab} - dX = 2 F_I F_I + K .}
Note that $L$ vanishes in the sectors $(0,4)$ and $(1,3)$ and in the
sector $(2,2)$ its flat components have the following structure

$$L_{ba\a\b} = (\g_{[a})_{\a\g}(\g_{b]})_{\b\d} L^{\g\d} ,$$ where

\eqn\lab{ L^{\a\b} = K^{\a\b} +  W_I^\a W_I^\b .} Moreover
$L^{\g\d}$ belongs to the $120$ representation of the Lorentz group
and therefore can be written as

\eqn\gabc{ L^{\b\g} = \g_{abc}^{\b\g} L^{abc}.}

The BRST variation of $\A$ will be of the form

\eqn\var { Q{\A} = \ha \int d^2z ~ \l^\a \l^\b \Pi^A \Pb^B
L_{BA\a\b} + \cdots . }

In the case of the GS heterotic string the analogous of  \anm\
(without $\cdots$) is the all story. Indeed in this case, the
anomaly and its variation are still given by \anm\ and \var\ but
with $\l^\a$ replaced by $\d \kappa_\g \Pi^c \g_{c}^{\g\a}$. With
this substitution  \var\ (without $\cdots$) vanishes (modulo the
Virasoro constraint) and \anm\ (without $\cdots$) is the full
consistency anomaly.

For the pure spinor string one must supplement the first term in the
r.h.s. of \anm\ with further terms (represented by the $\cdots$ ) in
order to recover a consistent anomaly. We shall show that the pure
spinor BRST anomaly is

\eqn\expec{ {\A} = \ha \int d^2z ~ [\l^\a \Pi^A \Pb^B \wh_{BA\a} -
\l^\a d_\b \Pb^a (\g_a)_{\a\g} L^{\b\g} - \l^\a \l^\b \o_\g \Pb^a
(\g_a)_{\b\r} \N_\a L^{\g\r}] , } For that we must compute the BRST
variation of ${\A} $ and show that $Q{\A}$ vanishes. The relevant
BRST transformations are \ChandiaIX\

\eqn\brst{ Q \Pi^A = \d^A_\a \N \l^\a - \l^\a \Pi^B
T_{B\a}{}^A,\quad Q \l^\a = 0 ,}
$$
Q d_\a = \l^\b \Pi^a (\g_a)_{\b\a} + \l^\b \l^\g \o_\d
R_{\a\b\g}{}^\d ,$$ and \BerkovitsFE, \ref\OdaSD{I.~Oda and
M.~Tonin, ``Y-formalism in Pure Spinor Quantization of
Superstrings,'' Nucl.\ Phys.\  B727 (2005) 176
[arXiv:hep-th/0505277].}

\eqn\brstdue{ Q \o_\a = d_\g ( \d^\g_\a - {\cal K}^\g{}_\a ) ,}
where

$$
{\cal K}^\g{}_\a = \ha (\g^aY)^\g (\l\g_a)_\a $$ and $Y_\a =
{{v_\a}\over {(v \l)}}$ so that $(Y\l)= 1$, $v_\a$ being a constant
spinor . Note that although we have added a non covariant object,
namely ${\cal K}$, the final result is covariant. Now it will be
shown that the anomaly \expec\ is invariant under the symmetry
transformation $\d \o_\a = (\g_a \l)_\a \L^a$. Then, the term $d_\g
{\cal K}^\g{}_\a$ in \brstdue\ does not contributes and \brstdue\
can be replaced by $Q \o_\a = d_\a$. To prove this consider first
the gauge part in \expec. After using $\N_\a W_I^\b = U_{I\a}{}^\g$,
we obtain

$$
\N_\a ( W_I^\b W_I^\g ) = U_I \d_\a^{[\b} W_I^{\g]} + {1\over 4}
U_I{}^{ab} ( \g_{ab} )_\a{}^{[\b} W_I^{\g]} ,$$ then plugging this
into the variation under $\d \o_\a = (\g_a \l)_\a \L^a$ we find that
${\A}$ varies as the integral of

$$
\L^b \Pb^a \l^\a \l^\b \l^\g (\g_b)_{\g\s} (\g_a)_{\b\r} ~ [ U_I
\d_\a^{[\s} W_I^{\r]} + {1\over 4} (\g_{cd})_\a{}^{[\s} W_I^{\r]}
U_I^{cd} ]$$
$$
= \L^b \Pb^a \l^\a \l^\b \l^\g ~ [ (\g_{[a})_{\a\b} (\g_{b]})_{\g\r}
U_I W_I^\r - {1\over 4} ( \g_{cd} \g_{[a} )_{\a\b} (\g_{b]} )_{\g\r}
U_I^{cd} W_I^\r ],$$ which vanishes because the pure spinor
condition.

Similarly, for the $K$-part we need to use the result \BonoraXN\ and
the mappings \redefvielbein, \redcon. We first define $K^{\b\g} =
\g_{abc}^{\b\g} K^{abc}$ to obtain

$$
\N_\a K^{\b\g} = (\g^{abc})^{\b\g} \N'_\a K_{abc} + 2 \Os_\a
K^{\b\g} - (\g^{abc})^{\b\g} (\g_a{}^d)_\a{}^\r K_{bcd} \Os_\r,$$
where
$$
\N'_\a K_{abc} = (\g_{[a})_{\a\b} K_{bc]}{}^\b,$$ as it was shown in
\BonoraXN. Plugging this into the variation of \expec\ under the
pure spinor gauge transformation, we obtain that the variation of
\expec\ becomes the integral of

$$
\L^b \Pb^a \l^\a \l^\b \l^\g (\g_{abcde})_{(\a\b} \g^c_{\g)\d} (
K^{de\d} + \g_f^{\d\r} K^{def} \Os_\r ),$$ which vanishes because of
the  identity

$$
(\g_{abcde})_{(\a\b} \g^c_{\g)\r} = - \ha \g^c_{(\a\b} ( \g_{abde}
\g_c )_{\g)\r} ,$$ and the pure spinor constraint.

Now let us compute  $Q {\A}$. It is not difficult to obtain, after
using \var, that

\eqn\qa{ Q {\A} = \ha \int d^2z ~( \l^\a \l^\b d_\g \Pb^a [ -
T_{a\a}{}^\b (\g_b)_{\b\r} L^{\g\r} + \N_\a ( (\g_a)_{\b\r} L^{\g\r}
) - (\g_a)_{\a\r} \N_\a L^{\g\r} ] }
$$
+ \l^\a \l^\b \l^\g \o_\d \Pb^a [ T_{a\a}{}^b (\g_b)_{\b\r} \N_\g
L^{\d\r} - \N_\a ( (\g_a)_{\b\r} \N_\g L^{\d\r} ) + R_{\r\a\b}{}^\d
(\g_a)_{\g\s} L^{\r\s} ]).
$$
If note that $\N_\a (\g_a)_{\b\g} = - 2 \Os_\a (\g_a)_{\b\g}$, the
Fierz identity for the gamma matrices and the pure spinor condition,
then the first line in \qa\ vanishes and we are left with the
expression from the last line that contains

\eqn\zero{ \l^\a \l^\b \l^\g [ R_{\r\a\b}{}^\d L^{\r\s} + \N_\a
\N_\b L^{\d\s} ] (\g_a)_{\g\s} .} If we symmetrize in $(\a\b)$, use

$$
\{ \N_\a , \N_\b \} L^{\d\s} = -T_{\a\b}{}^A \N_A L^{\d\s} +
L^{\r\s} R_{\a\b\r}{}^\d + L^{\d\r} R_{\a\b\r}{}^\s ,$$ and the
Bianchi identity $R_{(\a\b\r)}{}^\d = 0$, then we obtain that \zero\
is proportional to

\eqn\one{ \l^\a \l^\b \l^\g R_{\a\b\r}{}^\s (\g_a)_{\g\s} .} But

$$
R_{\a\b\r}{}^\s = \d^\s_\r R_{\a\b} + {1\over 4} (\g^{bc})_\r{}^\s
R_{\a\b bc} .$$ If we plug this expression into \one, we see that
$R_{\a\b}$ does not contribute because it contains a term like
$\g^c_{\a\b}$. Analogously $R_{\a\b bc}$ is expressed in terms of a
term along $\g^d_{\a\b}$, which again does not contribute, and a
term along $\g_{bcdef}$. We note that this contribution also
vanishes because of the identity

$$
(\g_a \g^{bc})_{\r(\a} (\g_{bcdef})_{\b\g)} = (\g_a \g^b)_\r{}^\s
\g^c _{\s(\a} ( \g_{bcdef} )_{\b\g)} = \ha (\g_a \g^b)_\r{}^\s (\g_c
\g_{bdef})_{\s(\a} \g^c_{\b\g)} .$$ Therefore we have obtained

\eqn\qqaa{ Q {\A} = 0. }

The anomaly ${\A}$ in \expec\ can be absorbed by relaxing the
torsion and curvature constraints that follow from \nilp\ and the
vanishing of \qaction\ and by modifying them. In fact we can impose

\eqn\qseff{ Q S - \a' {\A} = {1\over\a'} \int d^2 z [\ha \l^\a \Pi^a
\Pb^b ( T_{\a(ab)} + \H_{ba\a} ) + \ha \l^\a \Pi^\b \Pb^a ( \H_{\b\a
a} - T_{\b\a a} ) }
$$
 + \l^\a \Pi^a \Jb^I ( \ha( \H_{\a\b a} + T_{\a\b a} ) W^\b_I - F_{Ia\a} ) - \l^\a \Pi^\b
\Jb^I ( W_I^\g \H_{\g\a\b} +   F_{I\a\b} ) $$
$$
  + \l^\a
d_\b \Jb^I ( U_{I\a}{}^\b - W_I^\g T_{\g\a}{}^\b - \N_\a W_I^\b ) +
\l^\a \l^\b \o_\g ( \N_\a U_{I\a}{}^\g + W_I^\d R_{\d\a\b}{}^\g )
$$
$$
+ \l^\a d_\b \Pb^a ( T_{a\a}{}^\b - {\a'\over 2} (\g_a)_{\a\g}
L^{\b\g} ) - \l^\a \l^\b \Pb^a ( R_{a\a\b}{}^\g - {\a'\over 2}
(\g_a)_{\d(\a} \N_{\b)} L^{\g\d} )] = 0 ,$$
where we have defined \eqn\newh{ \H = dB - {\a'\over 2} \wh .}

Equation \qseff\ means that the structure of the anomaly is such
that a violation of the BRST invariance of the classical action $S$,
represented by a change of the constraints, can be chosen so that it
cancels the anomaly, as in the GS mechanism.

It follows from \qseff\ that the constraints
$$  T_{\a\b}{}^\g = 0,\quad T_{\a\b}{}^a =
\g^a_{\a\b},\quad T_{\a a}{}^b = 2(\g_a{}^b)_\a{}^\b \O_\b \quad
F_{I\a\b}= 0 ,$$ remain the same. Only the constraints \curv\
are changed in the sense that it is $\H$ and not $H$ that satisfies
these constraints. All the other components of the torsion and
curvatures follow from the Bianchi identities. In particular

$$T_{a\a}{}^\b = {\a'\over 2} (\g_a) _{\a\g} L^{\b\g} ,$$
and

$$ \l^\a \l^\b R_{a\a\b}{}^\g = \a' \l^\a \l^\b (\g_a)_{\d\a}
\N_\b L^{\g\d} ,$$ in agreement of the last two terms of \qseff.

We expect that the corrections we have found will induce a
correction in the nilpotence of the BRST charge, at the one-loop
level, that consists in replacing $H$ with $\H$ in the second
constraint in \nilp.

\newsec{Concluding Remarks}

In this paper we have obtained the corrections of order ${\a'}$ for
the constraints of the $\sigma$-model of the pure spinor heterotic
string, that implement the GS anomaly cancelation mechanism. They
arise as anomalies of the BRST charge. In fact, having worked at a
cohomological level, we have obtained the general form of these
corrections, which depends on two unspecified constants: one in
front of the gauge anomaly and one in front of the Lorentz one.
These constants are fixed as in \expec\ by requiring that the
variations of $B$ under gauge and Lorentz transformations, induced
by this BRST anomaly, cancel the gauge and Lorentz anomalies \local,
according to the GS mechanism. It could be interesting to check
these values of the constants by an explicit one loop calculation.

We have obtained our result in the framework of the set of
constraints found in \BerkovitsUE\ starting from \action. A
redefinition of the supervielbeins and superconnections leads to a
different but equivalent set of constraints. Of course the
redefinition changes the $\sigma$-model action \action\ but the new
action is equally suited and gives rise to equivalent results. For
instance the redefinitions \redefvielbein\ and \redcon\ lead to the
action

$$ S = {1\over\a'} \int d^2z ~[ \ha  e^{{2\over
3}\Phi} \Pi^a \Pb^b \eta_{ab} + \ha \Pi^A \Pb^B B_{BA} + \Pi^A
A_{IA} \Jb^I + \o_\a \Nb \l^\a
$$ $$
 + d_\a ( \Pb^\a + \Jb^I W_I^\a - {1\over 3} \Pb^a \g_a^{\a\b} \N_\b \Phi )
+ \l^\a \o_\b ( \Jb^I U_{I\a}{}^\b + {1\over 4} \Pb^A \L_A{}^{ab}
(\g_{ab})_\a{}^\b ) ] + S_J + S_{FT} ,$$ where $\L^{ab}$ is the one
form \redom\ expressed in terms of $E'^A$ (in this equation the
suffix `` ' '' is suppressed).

Notice that a change in the action $S$ not only changes the
constraints coming from the vanishing of \qaction\ but also induces
changes in the definition of $d_\a$ and therefore gives rise to
possible changes in the nilpotence motivated constraints \nilp. Also
notice that the anomaly $\A$ is defined modulo a trivial cocycle
that amounts to a modification of the action corresponding to an
($\a'$-dependent) redefinition of supervielbeins, B-superform and
superconnections \ref\LechnerXV{K.~Lechner `` String kappa Anomalies
and D = 10 Supergravity Constraints: The Solution of a puzzle,''
Phys.\ Lett.\ B357 (1995) 57.}.

In \ref\LechnerXI{A.~Candiello and K.~Lechner `` Duality in
Supergravity Theories''.\ Nucl.\ Phys.\ B412 (1994) 479.},
\LechnerXV, (see also \ref\Pasti{L.~Bonora,  P.~Pasti and M.~Tonin
``Chiral Anomalies in Higher Dimensional Supersymmetric Theories,''
\ Nucl. \ Phys.\ B286 (1987) 150.}) an interesting set of
constraints is proposed. For this set, the curvature $R^{ab}$ in the
sector $(0,2)$  vanishes at the classical level (zero order in
$\a'$) and the 3-superform $X$ is of order $\a'$ so that it does not
contribute in $\H$ at first order in $\a'$. Then \newh\ looks as $\H
= dB - {\a'\over 2} \ws - \a'^2 X$.

As it was shown in \BonoraXN, the explicit solution of the Bianchi
identities in the presence of the superform $X$ leads to an
unexpected and, at first sight, unpleasant feature: the solution
contains poles that represent spurious states of negative norm
(poltergheists) at a mass of the order ${\kappa\over \a'}$ where
$\kappa$ is the v.e.v. of the dilaton. The poltergheists are the
signal of a conflict between our requirements of supersymmetry,
locality and unitarity (absence of anomalies). However one should
not worry of them. Indeed the spurious poles arise at a very high
mass in a region of energy where our perturbative expansion in $\a'$
is clearly unreliable. This is similar, after all, to what happens
in the well-known low energy effective actions of gravity with terms
quadratic in the curvature, which also contain poltergheists. Notice
that in the set of constraints of \LechnerXI, \LechnerXV, $X$ does
not contribute at first order in $\a'$ and therefore the spurious
poles appear only at higher orders. Moreover the spurious states can
be decoupled at any finite order in $\a'$, at the expense of
locality, by solving recursively the relevant equations, as
discussed in \BonoraXN.

As a last remark, let us notice that the cohomological method of
this paper could be used to search for anomalies and corrections of
the constraints, for type II superstrings and/or for heterotic
strings at higher order in $\a'$. In particular it should be
interesting to search for the anomaly at the order $\a'^3$ that
corresponds to the celebrated ``$R^4 $'' term in the action and
would provide for the supersymmetrization of this term. Previous
attempts in this direction (for the heterotic GS string) are in
\ref\LechnerII{ L.~Lechner, P.~Pasti and M.~Tonin `` Anomaly Free
Sugra and the R**4 Superstring Term,'' \ Mod. \ Phys.\ Lett.\ A2
(1987), 929\semi K.~Lechner and P.~Pasti ``Nonminimal Anomaly Free D
= 10, N=1 Sugra-Sym and Four Graviton Superstring Amplitudes,'' Mod.
\ Phys.\ Lett.\ A4 (1989) 1721.}. Note that the complete $R^4$ terms
for the type II superstrings were obtained recently in
\ref\PolicastroVT{ G.~Policastro and D.~Tsimpis, ``R**4, Purified,''
Class.\ Quant.\ Grav.\  23 (2006) 4753 [arXiv:hep-th/0603165].} by
using tree-level scattering amplitudes in the pure spinor formalism.

\vskip 15pt {\bf Acknowledgements:} We would like to thank Kurt
Lechner, Ichiro Oda, Vladimir Pershin, Dmitri Sorokin and Brenno
Vallilo for useful comments and suggestions. We specially thank
Nathan Berkovits for very useful comments on the manuscript and for
very enlightening discussions on the subject of this paper. OC would
like to thank the Coimbra group for the fellowship under the
``Scholarships Programme for Young Professors and Researchers from
Latin American Universities,'' the Universit\`a di Padova for
hospitality; and the financial support from Fondecyt (grant
1061050), Universidad Andr\'es Bello (proyecto interno 27-05/R) and
the Fundaci\'on Andes. The work of MT was supported by the European
Community's Human Potential Programme under contract
MRTN-CT-2004-005104 ``Constituents, Fundamental Forces and
Symmetries of the Universe.''

\listrefs

\end